\documentclass[lettersize,journal]{IEEEtran}
\usepackage{amsmath,amsfonts}
\usepackage{algorithmic}
\usepackage{algorithm}
\usepackage{array}
\usepackage[caption=false,font=normalsize,labelfont=sf,textfont=sf]{subfig}
\usepackage{textcomp}
\usepackage{stfloats}
\usepackage[hidelinks]{hyperref}
\usepackage{url}
\usepackage{verbatim}
\usepackage{graphicx}
\usepackage{svg}
\usepackage{xcolor} 
\usepackage{cite}
\usepackage{listings}
\usepackage{minted}

\usepackage{tikzpagenodes}
\usepackage{eso-pic}

\begin{document}

\begin{center}
\small
\textit{This is the accepted version of the following article:}\\
B. Lopez, A. Diaz-Bricio, J. Perez, I. Vidal, and F. Valera, ``Quditto: Emulating and Orchestrating Distributed QKD Network
Deployments,''
\textit{IEEE Access}, 2026, DOI: \url{10.1109/MNET.2026.3736912}.
\end{center}

\vspace{0.5em}

\noindent
\textcopyright\ 2026 IEEE. Personal use of this material is permitted.  Permission from IEEE must be obtained for all other uses, in any current or future media, including reprinting/republishing this material for advertising or promotional purposes, creating new collective works, for resale or redistribution to servers or lists, or reuse of any copyrighted component of this work in other works.

\title{\textit{Quditto}: Emulating and Orchestrating Distributed QKD Network Deployments}

\author{Blanca Lopez, Angela Diaz-Bricio, Javier Perez, Ivan Vidal, and Francisco Valera }
 
\thanks{B. Lopez (email: bllopezr@it.uc3m.es) and A. Diaz-Bricio are with IMDEA Networks and the Telematics Department of Universidad Carlos III de Madrid. J. Perez, I. Vidal and F. Valera are with the Telematics Department of Universidad Carlos III de Madrid.}


\maketitle
\begingroup
\renewcommand\thefootnote{}
\footnotetext{

}
\endgroup

\begin{abstract}
Quantum Key Distribution (QKD) offers information-theoretic security by leveraging quantum mechanics, yet the cost and complexity of dedicated hardware and fiber infrastructure have so far limited large-scale deployment and experimentation. In this paper, we introduce \textit{Quditto}, an automated open-access emulation platform that combines configurable quantum-behavior modeling with a standardized key‐delivery API, enabling users to deploy selected QKD models as operational emulated QKD networks and interact with them exactly as they would with real QKD hardware. \textit{Quditto} modular design supports pluggable protocol implementations, complex key management schemes, and configurable channel models, including variable attenuation and decoherence. We validate \textit{Quditto} by deploying networks of various sizes and demonstrate its flexibility through two proof‐of‐concept scenarios featuring eavesdropper attacks and heterogeneous channel conditions.
\end{abstract}


\section{Introduction}
\IEEEPARstart{T}{he} progress of quantum computing offers promising advances across many areas, by allowing the solution of some complex computational problems at speeds far beyond the reach of classical algorithms. However, this great computational power threatens the cryptographic foundations underlying the security of current communication infrastructures. Many security protocols rely on mathematical problems, such as integer factorization and discrete logarithms, which could be solved in polynomial time using quantum algorithms. Faced with this situation, two parallel but potentially complementary paths open up to solve this security gap in Internet communications: post-quantum cryptography (PQC) and quantum cryptography, whose best-known method is Quantum Key Distribution (QKD).

PQC seeks to design new (non-quantum) cryptographic algorithms that are resistant to known quantum attacks, though it cannot guarantee that a future quantum algorithm will not eventually compromise their security. In contrast, quantum cryptography, particularly QKD protocols, leverages the intrinsic properties of quantum mechanics to achieve information-theoretic security. 

The ideal level of security offered by QKD protocols is, however, far from widespread deployment. First, the underlying hardware remains at an early stage of development, and it is delicate and costly. Second, quantum signals are highly susceptible to loss and decoherence over standard telecommunication fibers, so they must travel through dedicated, low-loss channels, which increases infrastructure expenses. As a result, building large-scale QKD Networks (QKDNs) is technically challenging and economically unfeasible for both commercial providers and most research institutions. 

For this reason, numerous simulation tools have been developed in recent years, allowing experimentation with objects representing QKDNs without the need to physically build one. Some examples, such as NetSquid \cite{netsquid}, allow detailed simulation of devices, channels, and protocols; while others, such as SimQN \cite{simqn}, focus on simulating what would be the network layer of a QKDN. However, to the best of our knowledge, none of these tools allow true end-to-end emulation of a distributed QKDN infrastructure (across distant nodes).

In this context, we previously built a service for automatically deploying Digital Twins of QKDNs \cite{qd2v1} as part of the Madrid quantum communication infrastructure development. While this approach enabled distributed emulated experimentation using SimulaQron \cite{simulaqron}, its quantum-behavior modeling lacked the precision required for detailed protocol- and channel-level performance evaluations.

Building on that groundwork, we now present \textit{Quditto} \cite{qd2}, a comprehensive QKD emulator that combines model-driven quantum-behavior characterization with an ETSI standard-compliant Application Programming Interface (API). The main contributions of this work are:
\begin{itemize}
    \item A system-level emulation architecture that decouples quantum-behavior modeling from node operation, enabling pluggable protocol backends.
    \item Automated, distributed deployment of emulated QKDNs across heterogeneous classical infrastructure via a single orchestration command, allowing a selected backend model to be instantiated as an operational multi-node network.
    \item The implementation of a standardized key-delivery API \cite{etsi_014} on all emulated nodes, enabling real-time interaction identical to that with physical QKD hardware.
    \item Validation through orchestration benchmarks and two proof-of-concept scenarios, showing how different backend models, including adversarial channel conditions and realistic photon-loss implementations, can be deployed and accessed through a distributed, standard-aligned QKDN.
\end{itemize}

This work is structured as follows: {Section II reviews the state of the art in the field of QKDN modeling and compares existing platforms with \textit{Quditto}; Sections III and IV deal respectively with the design and implementation of \textit{Quditto}; Section V presents the platform validation; finally, Section VI contains the conclusions of the work and future lines arising from it.

\section{State of the art}

\IEEEPARstart{D}{ue} to the existing difficulty in accessing quantum equipment, multiple simulation frameworks have been developed to support the design, analysis, and performance evaluation of QKDNs at different levels of abstraction.

At the device level, the leading platform is \textit{NetSquid} \cite{netsquid}, a discrete event simulator that enables modeling from single-photon generators to detectors and channels. These models can be used to create protocol implementations, analyzing them in simulations under realistic conditions. \textit{NetSquid} allows researchers to prototype new QKD schemes, but because it focuses on individual device interactions and channel physics, it does not natively support distributed network topologies or standardized interfaces for higher-level applications.

Network layer simulators have also been developed to explore aspects such as routing and key management in QKDNs. In \textit{SimQN} \cite{simqn}, low-level quantum dynamics are abstracted away to allow the user to focus on link establishment modeling, or scheduling policies. Similarly, \textit{SimulaQron} \cite{simulaqron} and \textit{QuNetSim} \cite{qunetsim} avoid detailed simulations of elementary quantum objects, facilitating the design and comparison of quantum network strategies at scale. These platforms allow users to examine metrics such as QKDN throughput or latency, although in some cases the lack of accurate models makes the results unrealistic. Furthermore, none of them expose an API layer that mimics real QKD devices; instead, users extract data dumps for offline analysis, without being allowed real-time interaction with the network. 

Other tools, such as \textit{QuISP} \cite{quisp} or \textit{QNE} \cite{qne}, offer users a more realistic experience. \textit{QuISP} simulates hybrid quantum and classical traffic on the same network, while \textit{QNE} enables real-time key exchanges within small topologies. However, these initiatives do not implement any standardized key-delivery API, making the user unable to experiment with the simulated networks as they would with a real deployment.

Our previous service \cite{qd2v1} introduced the distributed deployment capabilities for QKDNs, moving from simulation to emulation. However, its reliance on SimulaQron \cite{simulaqron} for backend modeling means that the quantum behavior of the network is not mirrored with sufficient accuracy. As a result, the experimental outcomes it produces are inadequate for realistic performance analysis of the emulated networks.

The absence of a standardized interaction between most of these platforms contrasts with the interface model defined for practical QKD systems, leaving a gap between research tooling and real deployments. In particular, ETSI GS QKD 014 \cite{etsi_014} defines a REST-based key-delivery API for QKD devices, specifying the request/response semantics that allow applications to retrieve cryptographic material from QKD nodes in an interoperable manner. By defining how applications request and obtain keys from QKD nodes, the standard provides a natural boundary between the QKD infrastructure and the classical applications that consume the generated keys.

{Overall, existing works offer diverse perspectives on the behavior of QKDNs, but none provide a complete emulation framework that simultaneously combines quantum modeling, multi-device topology support, and compatibility with the ETSI GS QKD 014 API \cite{etsi_014}, which together enable controlled and reproducible evaluations of QKDN operation. This gap motivates the development of a new research platform that combines all these features, allowing users to deploy an emulated, fully distributed, and standard-aligned QKDN.}

{
Building upon this analysis, \textit{Quditto} complements existing simulation frameworks, which either focus on accurate quantum-physical modeling or rely on physically simplified abstractions to enable scalable network-level studies. Instead of acting as a simulator at a specific abstraction level, \textit{Quditto} functions as a complete emulator enabling classical computing equipment to reproduce the functional and timing behavior prescribed by the selected QKD protocol model. Through its standards-compliant API, \textit{Quditto} supports real-time interactive and distributed deployments, faithfully reproducing the operational behavior of QKD infrastructure.}

\section{Design}

\IEEEPARstart{Q}{uditto} was designed to preserve the distributed emulation capability of the previous service \cite{qd2v1}, follow a standardized key-delivery API  for faithful QKD deployment reproduction, and provide a flexible framework for implementing and evaluating quantum network behavior. Its design emphasizes flexibility, enabling the modeling of different quantum technologies and protocols, while preserving accessibility through an intuitive interface and a modular architecture.

\begin{figure}[h!]
    \centering
    \includegraphics[width=\linewidth]{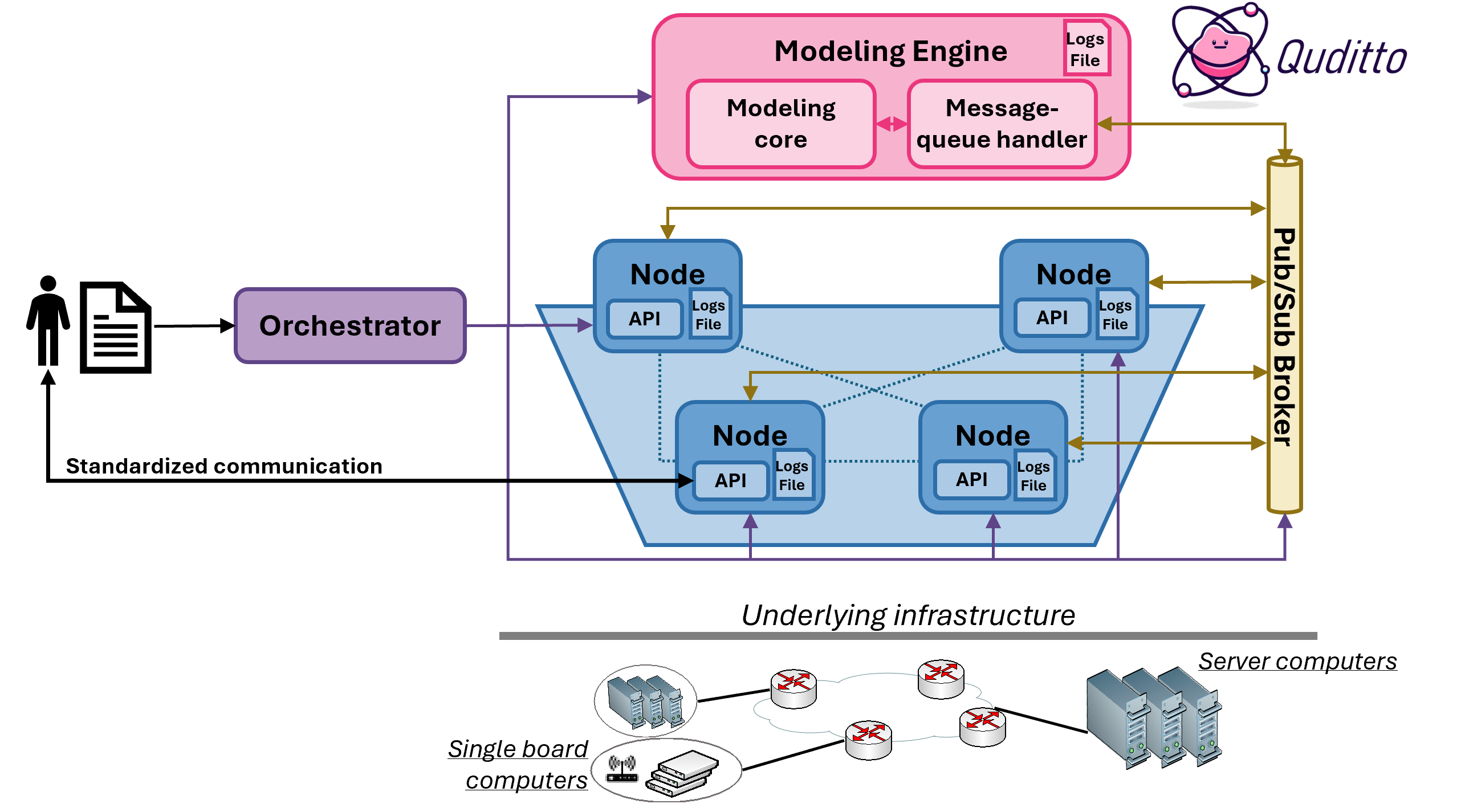}
    \caption{Outline of the general design of \textit{Quditto}. {The user composes a set of documents describing: 1) the desired QKDN, identifying its constituent nodes and the QKDN topology; 2) the parameters that define the behavior of each QKD link, such as distance, QKD protocol used for key exchange, or whether there is an eavesdropper; and 3) the set of equipment (standard PCs, single-board computers, virtual machines, or virtualization containers) that will be used to instantiate the QKD nodes. These documents are then passed to the \textit{Quditto} orchestrator. This component automatically installs and configures all required software on the target equipment. It then launches the \textit{Quditto} quantum-behavior modeling engine, which will be in charge of properly modeling the behavior of the QKD links, and instantiates lightweight \textit{Quditto} nodes on the target set of devices.}}
    \label{fig:general_design}
\end{figure}

\textit{Quditto} consists of three main components: the \textit{Quditto} orchestrator, the \textit{Quditto} modeling engine, and the \textit{Quditto} nodes. Platform usage should be straightforward and follow the design shown in Figure \ref{fig:general_design}.
From the user perspective, the resulting distributed QKDN behaves like a real QKDN: client applications communicate in real-time with network nodes via a standardized API, without ever needing to see the orchestration or modeling layers. Meanwhile, the modeling engine generates log files capturing events and protocol exchanges for monitoring, troubleshooting, and KPI measurement. 

\subsection{\textit{Quditto} Nodes}

The nodes handle application requests for cryptographic material. They offer a standardized API and internally forward each request to the modeling engine via a Pub/Sub protocol, using unique routing keys for message identification. When a client requests a key, the node publishes a “modeling request” message specifying its neighbor involved and the amount of cryptographic material needed. The modeling engine then executes the QKD protocol, generating the key and calculating the time it would have taken for physical equipment to create this key. Subsequently, it waits the specified time before publishing a “result” message tagged with the requesting node routing key, containing the key material. Upon the reception, the requesting node immediately returns the key to the client, reproducing the timing prescribed by the protocol model.
Likewise, the other node involved in the key exchange will receive its result message via its unique routing key (after the same calculated interval). This allows it to immediately answer any subsequent client queries once the interval has elapsed. 

In this way, \textit{Quditto} emulates a distributed QKDN that reproduces the timing and operational behavior of a physical deployment as prescribed by the selected protocol model, while remaining highly accessible and extensible. By supporting the core capability of key generation across emulated QKD links, \textit{Quditto} enables the incorporation of more complex key management schemes, including trusted node relays for end-to-end exchanges between non-adjacent network endpoints.

\subsection{\textit{Quditto} Modeling Engine}

The \textit{Quditto} modeling engine consists of two interrelated parts: a message‐broker handler, which pulls tasks and publishes results, and a modeling core, that executes quantum channel and key exchange protocol implementations. It operates as an asynchronous event-driven service, processing tasks for different QKD links in parallel but enforcing link-level serialization. That is, if multiple requests arrive for the same link, the engine respects the latency of the first protocol execution and adds it to the next. Once a protocol execution is finished, the message-broker handler waits for its associated time to pass and publishes the generated key material back to the broker under the routing key of the requesting node and its peer. 

Communication with the nodes follows a well-defined schema. Each modeling request carries the identifiers of the two endpoint nodes, the requested key length in bits, and a unique routing key for result delivery. Once the protocol execution completes, the message-broker handler enforces the computed protocol duration before publishing a modeling result containing the generated key material, its assigned
ID, and that duration, addressed to the routing keys of both participating nodes.

This separation between the broker handler and the modeling core ensures the flexibility of \textit{Quditto}. To integrate an alternative backend, the implementation only needs to accept the modeling parameters defined by the user, and return the generated key material together with the calculated execution time. This makes it straightforward to integrate custom protocol extensions or different modeling platforms, including implementations that incorporate complex mechanisms such as error reconciliation, privacy amplification, and key distillation.


\subsection{\textit{Quditto} Orchestrator}

The orchestrator takes as input the network description and handles its deployment and configuration. It automatically provisions each target device with the required software stack, and sets up the message broker that mediates communication between the nodes and the modeling engine. 

Once all dependencies are in place, the orchestrator launches the necessary processes for the correct functioning of the network, including the broker message handlers, the interface for client requests, and the modeling core, producing a fully operational, standard-compliant network without any manual steps beyond provisioning the Python-compliant devices and drafting the initial topology document.

\section{Implementation}

\begin{figure*}
    \centering
    \includegraphics[width=\linewidth]{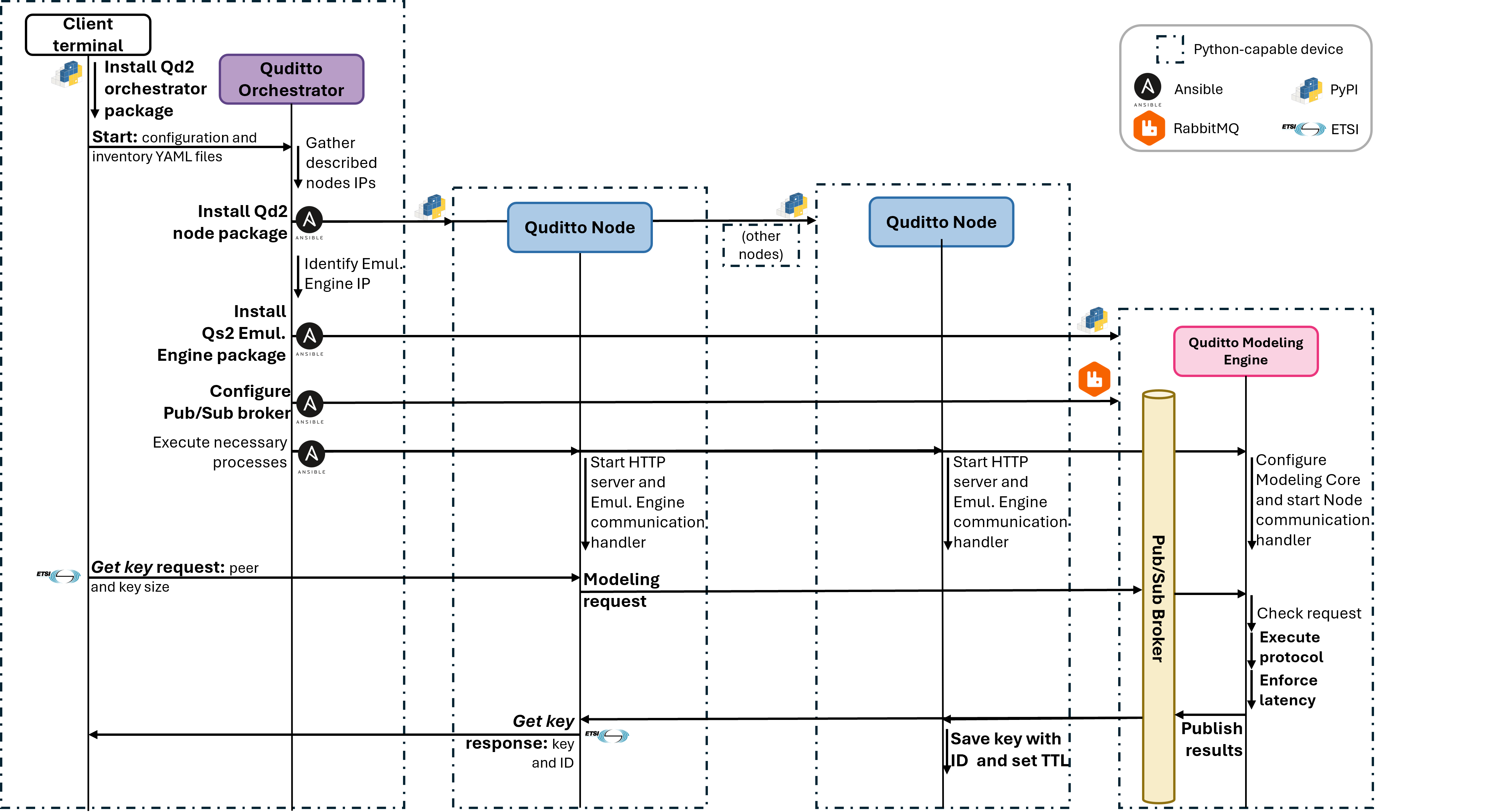}
    \caption{Flowchart of the instantiation of an emulated QKDN using \textit{Quditto}.}
    \label{fig:implementation}
\end{figure*}

\IEEEPARstart{Q}{uditto} is built as a Python-based, open-source platform \cite{qd2}. Each core component, namely the nodes, the modeling engine, and the orchestrator, is provided as an individual package. 

The orchestrator package uses Ansible \cite{ansible} to configure pre-existing devices with Internet connectivity, including desktop computers, single-board devices, virtual machines, or containers, as QKD nodes in the emulated network. The user must define the QKDN to be emulated, specifically, providing the \textit{Quditto} orchestrator with two YAML files, a commonly used language for configuration documents. The configuration file, describes the network topology, indicating node names, link lengths, possible eavesdroppers, and other parameters relevant to the QKD protocol to be emulated. The inventory file, includes the credentials used by Ansible to access the devices that will form the emulated QKDN.

Once both documents are provided, the orchestrator installs the node package on every node and the modeling engine package on the designated engine device. Therefore, the user only needs to install the orchestrator package, while the remaining installation and configuration steps are performed automatically. After the packages and Pub/Sub broker are configured, the orchestrator launches the processes reviewed in the previous section. The network becomes operational, allowing clients to issue real-time requests to the nodes through the API \cite{etsi_014}. 

When a client sends an initial request to a \textit{Quditto} node, the node generates a unique request ID and publishes a ``modeling request" under its routing key. After the emulation core completes the QKD protocol process and the calculated time has elapsed, it sends a ``modeling result" with the cryptographic material and its ID. The initiator node consumes this result and returns both the key and its ID to the client. The peer node also receives its own result through the broker, storing the key along with its ID in a local database for later retrieval. This complete flow is shown in Fig. \ref{fig:implementation}.

\subsection{\textit{Quditto} Node package}

To ensure efficient node operation, two separate processes running in parallel are implemented. 

The first process handles all interactions with the modeling engine performed through the broker implemented using the open-source RabbitMQ software \cite{rabbitmq}. The second hosts an HTTP server that provides the ETSI GS QKD 014 API \cite{etsi_014}. 

The implemented endpoints are \texttt{GET /api/v1/keys/enc\_keys} used by the session initiator to request new key material and, \texttt{GET /api/v1/keys/dec\_keys?key\_ID=\{key\_id\}} used by the responder to retrieve an existing key by its identifier. API conformance was verified by issuing requests to both a Quditto node and a real commercial QKD device and confirming that both returned structurally equivalent JSON responses matching the standard schema. This allows applications to make requests for cryptographic material using standardized primitives depending on their role in the communication. If the node receives a key request with an ID, it looks up that specific ID in its local store. If the ID is found, the node returns the associated cryptographic material without invoking the modeling engine. Keys are assigned a time-to-live of 10 minutes by default, following the design of commercial QKD devices; this value can be adjusted to suit application requirements.

\subsection{\textit{Quditto} modeling engine package}

The \textit{Quditto} modeling engine package is also comprised of two main components. The first one is responsible of communicating with the nodes via RabbitMQ. The second is the quantum-behavior modeling engine, which provides the quantum-level behavior associated with each emulated QKD link. By default, the \textit{Quditto} quantum-behavior modeling engine features {two QKD protocol NetSquid implementations based on the BB84 scheme: \textit{BB84 with Eve}, which enables the placement of an eavesdropper performing an intercept–resend attack at any point along the quantum channel, and \textit{Extended BB84} \cite{andres, andres2}, which incorporates user-configurable parameters to model the realistic behavior and imperfections of quantum hardware.} Additional scripts can be installed in \textit{Quditto} with alternative QKD protocol implementations, whether based on NetSquid or any other modeling platform, as long as they produce cryptographic material along with a execution duration.

When a ``modeling request" reaches the emulation core through the RabbitMQ broker, it checks that it is a legitimate request, i.e., that the nodes involved exist in the network and that they are neighbors. If so, it proceeds to execute the protocol implementation script as many times as necessary to collect the amount of cryptographic material needed to satisfy the request. Because the modeling engine actually tracks how many bits remain to satisfy each request, protocol implementations do not need to accept a key‐length parameter, simplifying the integration of new protocols. At the same time, this separation allows developers to adopt any key‐generation strategy they choose (for example, maintaining an internal buffer) to fulfill requests. Once all the required key material has been gathered, the modeling engine enforces the calculated latency before publishing a ``modeling result” message. That message, which includes the key and its associated ID, is sent under the routing keys of both participating \textit{Quditto} nodes.

Throughout the process, the modeling engine records logs of QKD protocol interactions, including timing events and quantum bit error rate (QBER) metrics. These logs are available to users for analysis and debugging, allowing in-depth inspection of protocol behavior, performance, and anomaly detection. 

\subsection{\textit{Quditto} orchestrator package}

The orchestrator is implemented in Python as a command-line interface application with two primary functions, \textit{start} and \textit{stop}. The first function receives the already mentioned configuration and inventory files as arguments. The second function requires only the inventory document.

With the \textit{start} function, the orchestrator uses the Ansible Python API to connect to each device designated as a node and install the \textit{Quditto} node package. It then connects to the device that will serve as the modeling engine and provides it with a network description file including the appropriate parameters for protocol modeling, such as link lengths or the presence of eavesdroppers. The orchestrator also installs NetSquid in this device, as it is the default modeling platform, in addition to the \textit{Quditto} modeling engine package and RabbitMQ. Lastly, it configures a RabbitMQ broker to allow communication between the modeling engine and nodes.

Once this is done, the orchestrator informs every node of the broker address so they know where to send modeling requests. Finally, it launches all required scripts on each device so that the emulated QKDN comes fully operational.

The \textit{stop} command shuts down the entire emulated QKDN by connecting to each device in the inventory and terminating all processes associated with the \textit{Quditto} modeling engine and node packages.

\section{Validation}
This section validates \textit{Quditto} as a platform for automatically deploying and configuring distributed emulated QKDNs, testing different protocol models. The YAML files used in the experiments are available at \cite{qd2} to support reproducibility.

The evaluation is not intended to calibrate Quditto against a specific commercial QKD device or to benchmark it directly against simulators operating at different abstraction levels. Commercial QKD systems typically rely on proprietary protocol implementations and undisclosed device parameters, which prevent reproducible numerical comparison. Similarly, tools such as NetSquid in isolation or SimQN expose different interfaces, abstraction levels, and output metrics, making direct comparison inconclusive. Quditto is not intended as a replacement for these tools, and is therefore evaluated as a platform that enables the distributed, API-accessible emulation of any QKD model implementation.

\subsection{Performance of Orchestration Actions}

We first evaluate how \textit{Quditto} is able to automatically build a distributed emulated QKDN, measuring the time required by the orchestrator to transform a set of generic virtual machines into a functional QKDN. This experiment validates one of \textit{Quditto}’s core contributions: the capacity to perform automated and scalable deployment of emulated QKDNs with minimal user effort.

Fig.~\ref{fig:prueba1} shows the time breakdown of the orchestration process as the number of nodes increases. 

\begin{figure}[h!]
    \centering
    \includegraphics[width=\linewidth]{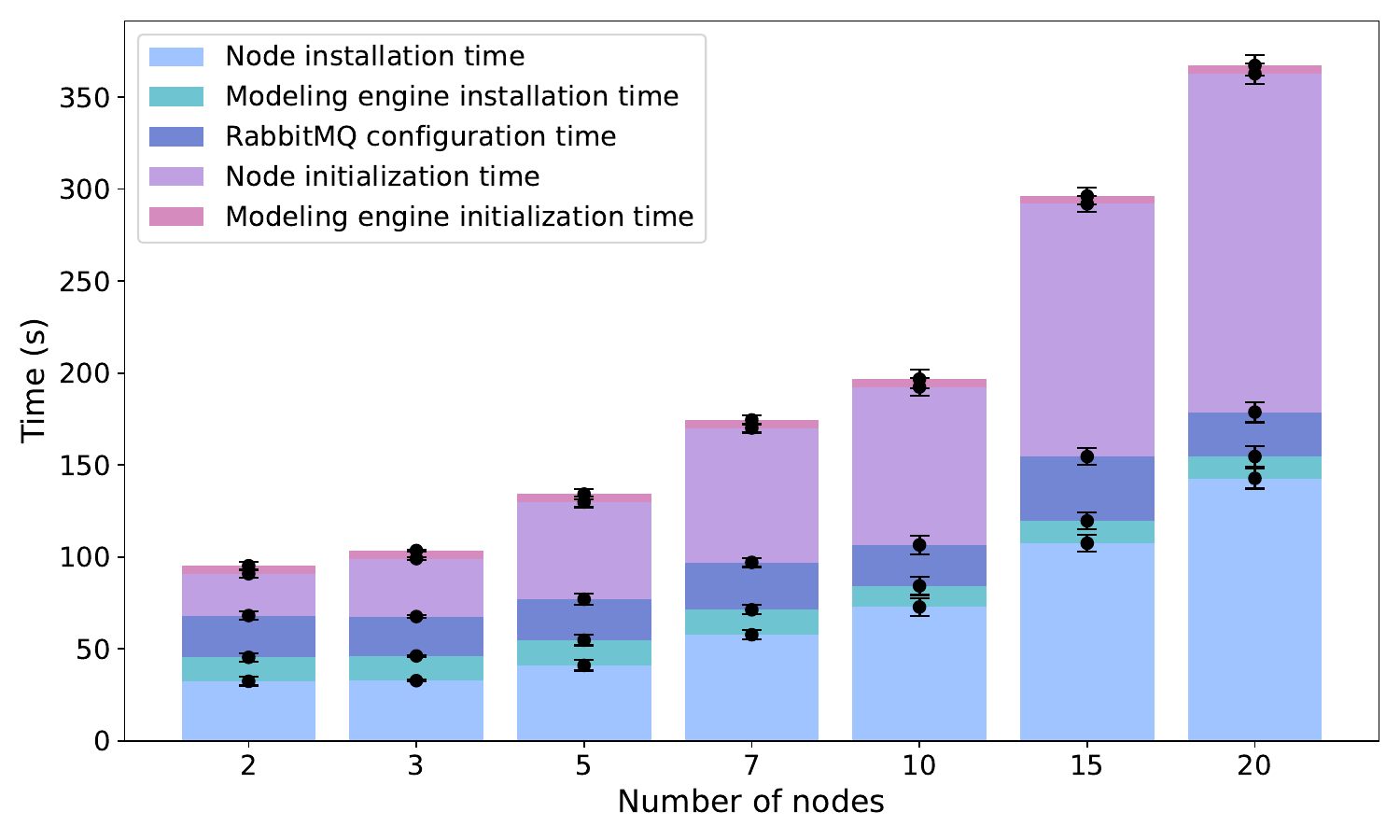}
    \caption{Detailed time of different actions  performed by\textit{Quditto} orchestrator. The distribution of 10 deployments with a 95\% confidence interval is shown for each network. The total setup time is decomposed into five distinct stages performed by the orchestrator in sequence via Ansible:
{(1) the node installation is the time taken to install the node package on all virtual machines specified in the configuration file, (2) the modeling engine installation includes both the installation of NetSquid and the modeling engine package on the designated modeling engine machine, (3) the RabbitMQ configuration represents the time required to install and configure RabbitMQ to enable communication between the nodes and the modeling engine, (4) the node initialization is the time it takes to run the required scripts on all nodes, (5) the modeling engine initialization corresponds to the time it takes to run the modeling engine scripts.
}}
    \label{fig:prueba1}
\end{figure}

Overall orchestration time increases with the network size. Fig.~\ref{fig:prueba1} shows that node installation and initialization grow linearly with the number of nodes (note that the separation in the horizontal axis does not scale linearly), since these actions are repeated for each node. In contrast, the modeling engine installation, RabbitMQ configuration, and modeling engine initialization times remain roughly constant, as they are performed only once on the modeling engine machine, regardless of the total number of nodes. {Therefore, node-related tasks dominate the orchestrator’s workload, while engine-specific steps add no additional overhead when the network grows.}

Remarkably, up to 10 nodes, \textit{Quditto} deploys faster (or similarly, in the case of 10) than even a 2-node network in our previous service \cite{qd2v1}. With 20 nodes, \textit{Quditto} deploys in less time than the previous system required for an 8-node network. This contrast highlights \textit{Quditto} efficiency gains despite its added functionality and complexity.

To assess modeling engine scalability under concurrent load, we tested three scenarios of increasing demand: 10 links with 10 simultaneous requests (100 total), 20 links with 10 requests each (200 total), and 10 links with 20 requests each (200 total). Completion times stayed below 60\,s, 80\,s, and 120\,s respectively, confirming that processing time scales linearly with load volume. Notably, per-request latency remained consistent across all scenarios, demonstrating that the modeling engine sustains throughput and accuracy without degradation as the number of concurrent requests grows.

\subsection{Performance in a Partially Adversarial QKD Environment}

\begin{figure}[b!]
    \centering
    \includegraphics[width=\linewidth]{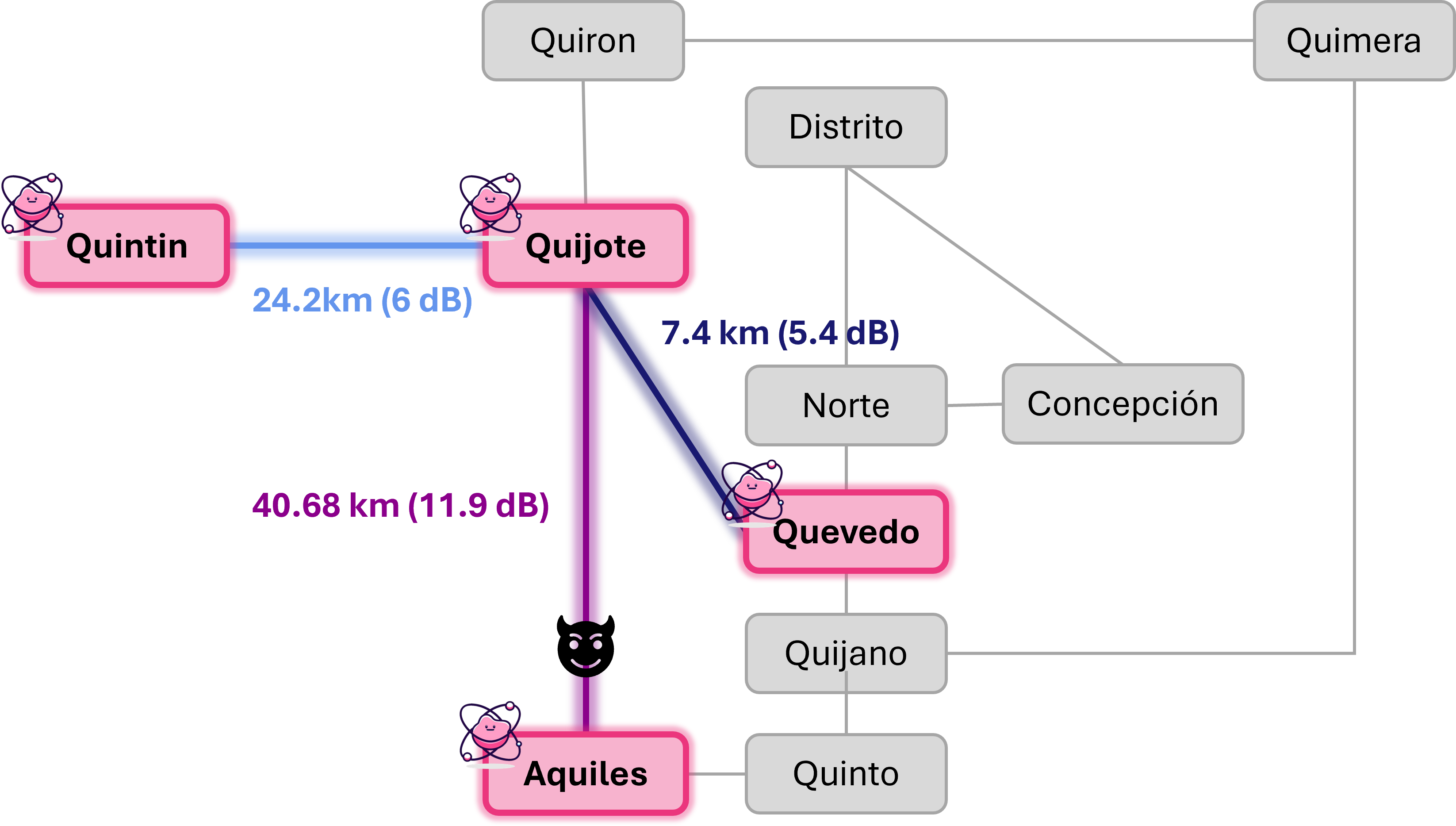}
    \caption{Madrid quantum communication infrastructure \cite{madq} with the nodes that we emulated for validation highlighted in color. {This topology is used as an illustrative example to demonstrate the platform’s operation}. Note that in Validation B, \textit{Performance in a Partially Adversarial QKD Environment}, the attenuations indicated in the figure were not taken into account, but we placed an eavesdropper on the link between \textit{Quijote} and \textit{Aquiles}. On the other hand, in Validation C, \textit{Performance in a Realistic QKD Environment}, we removed the eavesdropper and included the attenuated fibers.}
    \label{fig:red}
\end{figure}

\begin{figure*}[h!]
    \centering
    \includegraphics[width=\textwidth]{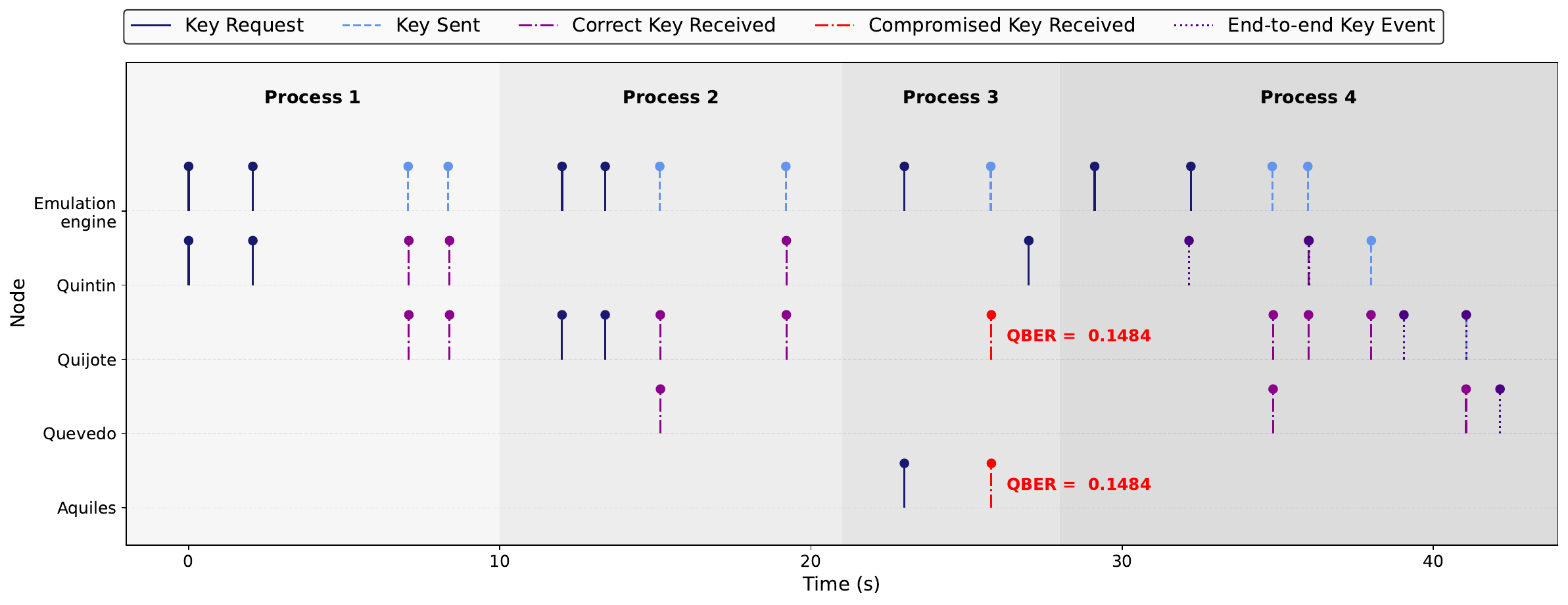}
    \caption{Events diagram with \textit{Quditto} nodes and modeling engine workflows.}
    \label{fig:prueba2}
\end{figure*}

{To functionally validate \textit{Quditto} and assess the logical sequence of events followed by the nodes and modeling engine during standard and adversarial key exchange workflows, we conducted a distributed experiment involving four nodes arranged in a simple star topology, reflecting a subset of the Madrid Quantum Infrastructure \cite{madq}.} The chosen portion of the network comprises nodes \textit{Quintin}, \textit{Quijote}, \textit{Quevedo}, and \textit{Aquiles} (see Figure~\ref{fig:red}). This setup captures node-engine interactions under controlled conditions, including the emulation of a limited adversarial scenario with an eavesdropper on the \textit{Quijote}–\textit{Aquiles} link. {The experiment employs one of the default QKD protocols available in \textit{Quditto}, \textit{BB84 with Eve}, which assumes ideal quantum channels while supporting the modeling of eavesdropping activity. 

Fig.~\ref{fig:prueba2} shows the four different processes that were evaluated. 

The first process evaluates two consecutive key requests on the same link. A user requests consecutively two keys via node \textit{Quintin} to the \textit{Quintin}–\textit{Quijote} link: first a long 512-bit key, followed by a 64-bit key. The modeling engine receives both requests, executes them, and delivers the final keys to the involved nodes. Receiving the longer key first confirms that \textit{Quditto} serializes multiple requests on the same link, completing the first request before starting the second.

The second process analyzes concurrent key requests on different links.  Two keys are requested via node \textit{Quijote}, one for the \textit{Quintin}–\textit{Quijote} link (512 bits) and another for the \textit{Quijote}–\textit{Quevedo} link (64 bits). Here, the shorter key (\textit{Quijote}–\textit{Quevedo}) arrives earlier, demonstrating \textit{Quditto} processes requests in parallel when they involve different links.

In the third process, a key is requested on the \textit{Quijote}–\textit{Aquiles} link, which is compromised by an eavesdropper intercepting 30\% of the qubits. This percentage was selected because it reliably places the QBER above the widely used BB84 security threshold of 0.11~\cite{bound}, making the compromised exchange unambiguous in the event diagram. The \textit{BB84 with Eve} implementation intentionally delivers the key material (different for each node) regardless of the 0.1484 QBER value, and records the metric in the modeling engine log. Threshold-based discard decisions are delegated to the calling application or key manager, in line with the separation of concerns between transport and policy layers; nevertheless, alternative protocol implementations could also enforce such discard policies directly at the protocol level.

Finally, the fourth process illustrates how \textit{Quditto} can flexibly support more complex configurations. In this case, we implemented a simple Key Management Entity (KME) capable of orchestrating a key relay via a trusted node. \textit{Quditto} enables this approach by supporting the  ETSI GS QKD 014 API \cite{etsi_014} for key requests, which allows external components like the KME to interact seamlessly with the nodes. The implemented scripts enable \textit{Quintin} and \textit{Quevedo}, which are not directly connected, to exchange a key by a relaying scheme with node \textit{Quijote} acting as a trusted node. Upon receiving a shared-key request with \textit{Quevedo}, \textit{Quintin} triggers two key requests to the engine: one for the \textit{Quintin}–\textit{Quijote} link and another for the \textit{Quijote}–\textit{Quevedo} link. After that, \textit{Quintin} generates the end-to-end key. Once the engine distributes the quantum keys to \textit{Quevedo} and \textit{Quijote}, and to \textit{Quintin} and \textit{Quijote}, \textit{Quintin} encrypts the end-to-end key with its shared key with \textit{Quijote}. These two actions happen in node \textit{Quintin} within milliseconds of each other and thus appear superposed in the diagram. Shortly after, \textit{Quintin} sends the encrypted end-to-end key to \textit{Quijote}, which decrypts it using the key shared with \textit{Quintin}. After that, \textit{Quijote} re-encrypts it using the key shared with \textit{Quevedo}, and forwards it to \textit{Quevedo}. These two events also appear nearly simultaneous. Finally, \textit{Quevedo} receives and decrypts the key, recovering the original end-to-end key created by \textit{Quintin}. As a result, \textit{Quintin} and \textit{Quevedo} share a secure key.

\subsection{Performance in a Realistic QKD Environment}
{Finally, the third experiment demonstrates \textit{Quditto} flexibility in suppongting different QKD protocol implementations, using \textit{Extended BB84} as an example. This model represents the quantum channel in more detail by incorporating channel attenuation, photon loss, and decoherence, detector inefficiencies and dark counts. The experiment uses the same network topology shown in Fig.~\ref{fig:red}; but omits the eavesdropper, and includes fibre photon losses, illustrating \textit{Quditto}’s ability to model more physically realistic scenarios. All other physical parameters are fixed to the default values of the hardware-informed model in \cite{andres}, which provides experimentally motivated values for channel attenuation, detector efficiencies, and dark count rates widely reported in the QKD literature.

\begin{figure}[!]
    \centering
    \includegraphics[width=\linewidth]{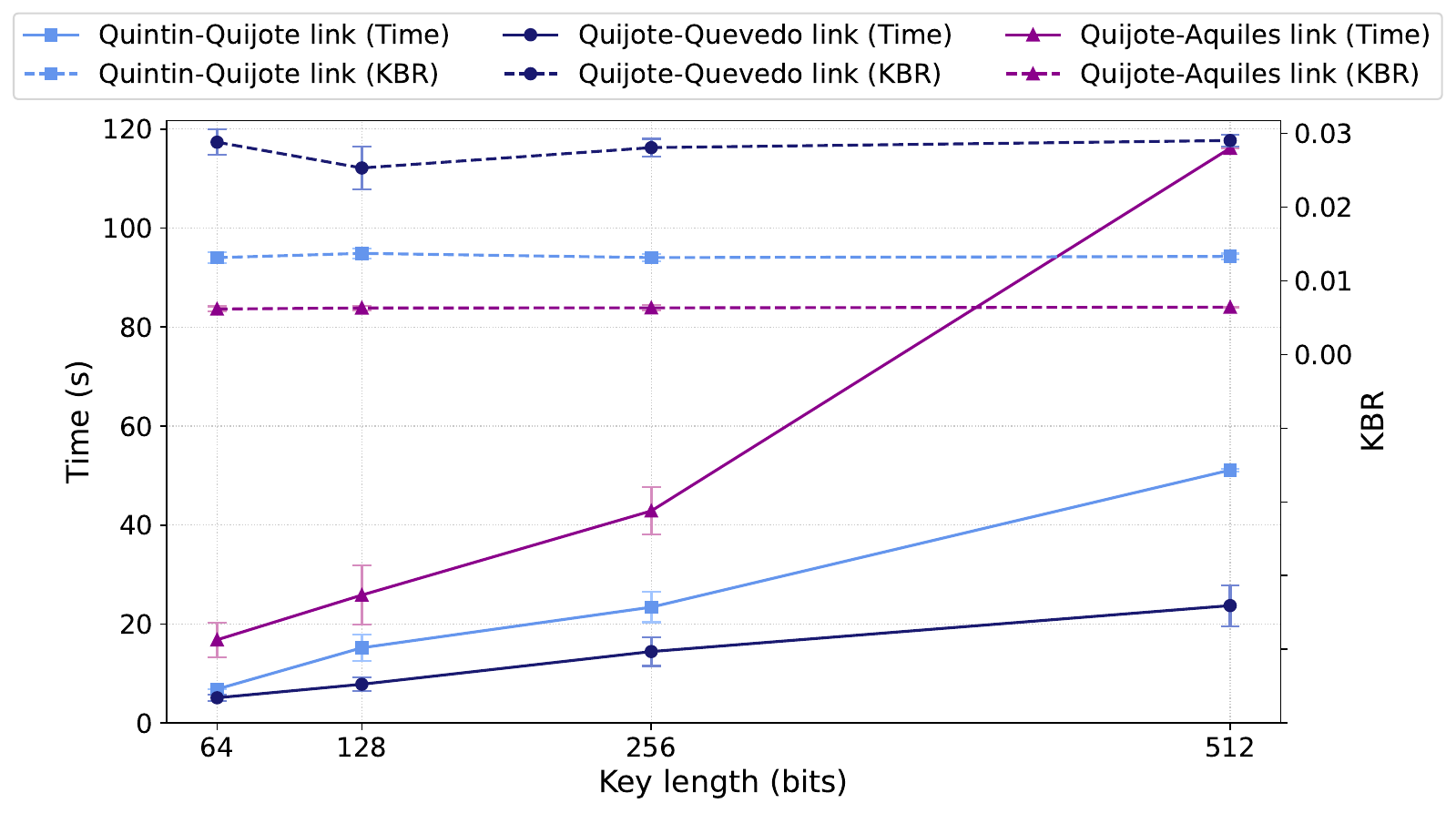}
    \caption{Performance of \textit{Quditto} in terms of time (left axis) and KBR (right axis) under realistic conditions, shown as a function of key size. For each key size, the key exchange was performed 10 times to gather statistics, and a 95\% confidence interval was calculated.}
    \label{fig:prueba3}
\end{figure}

Fig.~\ref{fig:prueba3} analyzes the average key exchange time and {Key Bit Rate (KBR)} for the three network links as key length increases.

On the left axis, key distribution time grows with the key length in all links, as expected. However, slope and absolute values depend on link characteristics. The \textit{Quijote}-\textit{Aquiles} link exhibits the highest times and the steepest growth rate. This is coherent with the physical properties of the link, which is the longest (40.68 km) and suffers the highest attenuation (11.9 dB). On the contrary, the \textit{Quijote}-\textit{Quevedo} link is the least lossy (5.4 dB) and the shortest one (7.4 km), which makes it the most time-efficient.

On the right axis, KBR, {captures the number of output bits per channel use,} measuring quantum channel efficiency. All links display relatively stable KBR values across key lengths, which is expected since KBR primarily depends on the intrinsic properties of the quantum channel {and should not vary with the key length}. Among them, the \textit{Quijote}-\textit{Quevedo} link achieves the highest KBR, consistent with its shorter distance and lower attenuation. This trend is inverse to the time measurements: links with better physical properties generate keys faster and yield higher KBRs, confirming the strong impact of channel quality on overall system performance. 

\section{Conclusion}

In this paper, we have presented \textit{Quditto}, a comprehensive emulation platform that combines configurable QKD model execution with support for multi-device topologies, and the implementation of the ETSI GS QKD 014 API \cite{etsi_014} as real-time communication mechanism. Its modular design allows developers to integrate different implementations of quantum devices, protocols, or management features. 

We have validated \textit{Quditto}, gathering metrics of its performance, and conducted proofs of concept in which we swapped out the emulation core to test different protocol implementations, including eavesdroppers and different channel models.

Three other main lines of research emerge from this project. First, incorporating additional QKD protocols, along with their reconciliation and privacy amplification modules, to expand the range of experiments. Second, extending the modeling engine itself to support its distributed deployment across multiple machines (horizontal scaling), further increasing the number of concurrently active links it can support. Third, supporting hybrid deployments by directly interfacing with quantum hardware for hardware-in-the-loop testing.

\section*{Acknowledgments}

This work has been partially funded by the project 6GINSPIRE PID2022-137329OB-C42, funded by MCIN/AEI/10.13039/501100011033/. This publication is part of the I+D+I project MADQuantum-CM, financed by the European Union NextGeneration-EU, Madrid Government and by the PRTR. This work has also been partially supported by the EU Horizon Europe project Quantum Security Networks Partnership (QSNP), under grant 101114043.
 

\begin{thebibliography}{1}
\bibliographystyle{IEEEtran}

\bibitem{netsquid}
Coopmans, T.; Knegjens, R.; Dahlberg, A.; Maier, D.; Nijsten, L.;
  de~Oliveira~Filho, J.; Papendrecht, M.; Rabbie, J.; Rozpedek, F.; Skrzypczyk,
  M.;  et~al.
\newblock NetSquid, a discrete-event simulation platform for quantum networks.
\newblock {\em Commun. Phys.} {\bf 2021}, \emph{4}, 164.
  \href{http://arxiv.org/abs/2010.12535}{{\normalfont
  [arXiv:quant-ph/2010.12535]}}.
\newblock {\url{https://doi.org/10.1038/s42005-021-00647-8}}.

\bibitem{simqn}
L. Chen et al., 
\newblock"SimQN: A Network-Layer Simulator for the Quantum Network Investigation," 
\newblock in IEEE Network, vol. 37, no. 5, pp. 182-189, Sept. 2023, 
\newblock doi: 10.1109/MNET.130.2200481.

\bibitem{qd2v1}
Martin, R.; Lopez, B.; Vidal, I.; Valera, F.; Nogales B. 
\newblock Service for Deploying Digital Twins of QKD Networks. 
\newblock Applied Sciences. 2024; 14(3):1018. 
\newblock \url{https://doi.org/10.3390/app14031018} 

\bibitem{simulaqron}
Dahlberg, A.; Wehner, S.
\newblock SimulaQron---A simulator for developing quantum internet software.
\newblock {\em Quantum Sci. Technol.} {\bf 2018}, {\em 4},~015001. 
\newblock {\url{https://doi.org/10.1088/2058-9565/aad56e}}.

\bibitem{qd2}
Lopez, B.; Diaz-Bricio, A.; Perez, J.; Vidal, I.; Valera, F.
\newblock
\emph{\textit{Quditto} repository} \newblock
 \url{www.quditto.io}

\bibitem{etsi_014}
ETSI, \emph{Quantum Key Distribution (QKD); Protocol and Data Format of REST-Based Key
  Delivery API}; ETSI: Sophia Antipolis, France, 2019.

\bibitem{qunetsim}
Diadamo, S.; Nötzel, J.; Zanger, B.; Beşe, M.M.
\newblock QuNetSim: A Software Framework for Quantum Networks.
\newblock {\em IEEE Trans. Quantum Eng.} {\bf 2021}, {\em
  2},~1--12.
\newblock {\url{https://doi.org/10.1109/TQE.2021.3092395}}.

\bibitem{quisp}
Satoh, R.; Hajdušek, M.; Benchasattabuse, N.; Nagayama, S.; Teramoto, K.; Matsuo, T.; Metwalli, S.A.; Satoh, T.; Suzuki, S.; Meter, R.V. 
\newblock QuISP: a Quantum Internet Simulation Package. 
\newblock 2022 IEEE International Conference on Quantum Computing and Engineering (QCE), 353-364.

\bibitem{qne}
Wehner, S.; Seidler, N.; Karpat, Ö.; et al. 
\newblock
\emph{Quantum Network Explorer (QNE)} \newblock
 \url{www.quantum-network.com}

\bibitem{ansible}
Ansible Community Documentation.
\newblock Available online: \url{https://docs.ansible.com}

\bibitem{rabbitmq}
RabbitMQ
\newblock Available online: \url{https://www.rabbitmq.com}

\bibitem{madq}
Lopez, D.; Brito, J. P.;  Pastor, A.; Martin, V.; Sánchez, C.
\newblock Madrid Quantum Communication Infrastructure: a testbed for assessing QKD technologies into real production networks 
\newblock 2021 Optical Fiber Communications Conference and Exhibition (OFC), San Francisco, CA, USA, 2021, pp. 1-4.

\bibitem{bound}
Scarani, V.; Bechmann-Pasquinucci, H.; Cerf, N.J.; Dušek, M.; Lütkenhaus, N.; Peev, M.
\newblock The Security of Practical Quantum Key Distribution
\newblock Rev. Mod. Phys. 2009, 81, 1301–1350.

\bibitem{andres}
Martin-Megino A.
\newblock
\emph{\textit Non-ideal-QKDNs} \newblock Available online:
 \url{https://github.com/amartin-m/Non-ideal-QKDNs}

\bibitem{andres2} 
Martin-Megino, A.; Lopez, B.; Vidal, I.; Valera, F.
Efficient QKD in Non-Ideal Scenarios with User-Defined Output Length Requirements.
 Technologies (2026), 14(5), 284. \url{https://doi.org/10.3390/technologies14050284}


\end{thebibliography}
%

\vfill



\begin{IEEEbiography}[{\includegraphics[width=1in,height=1.25in,clip,keepaspectratio]{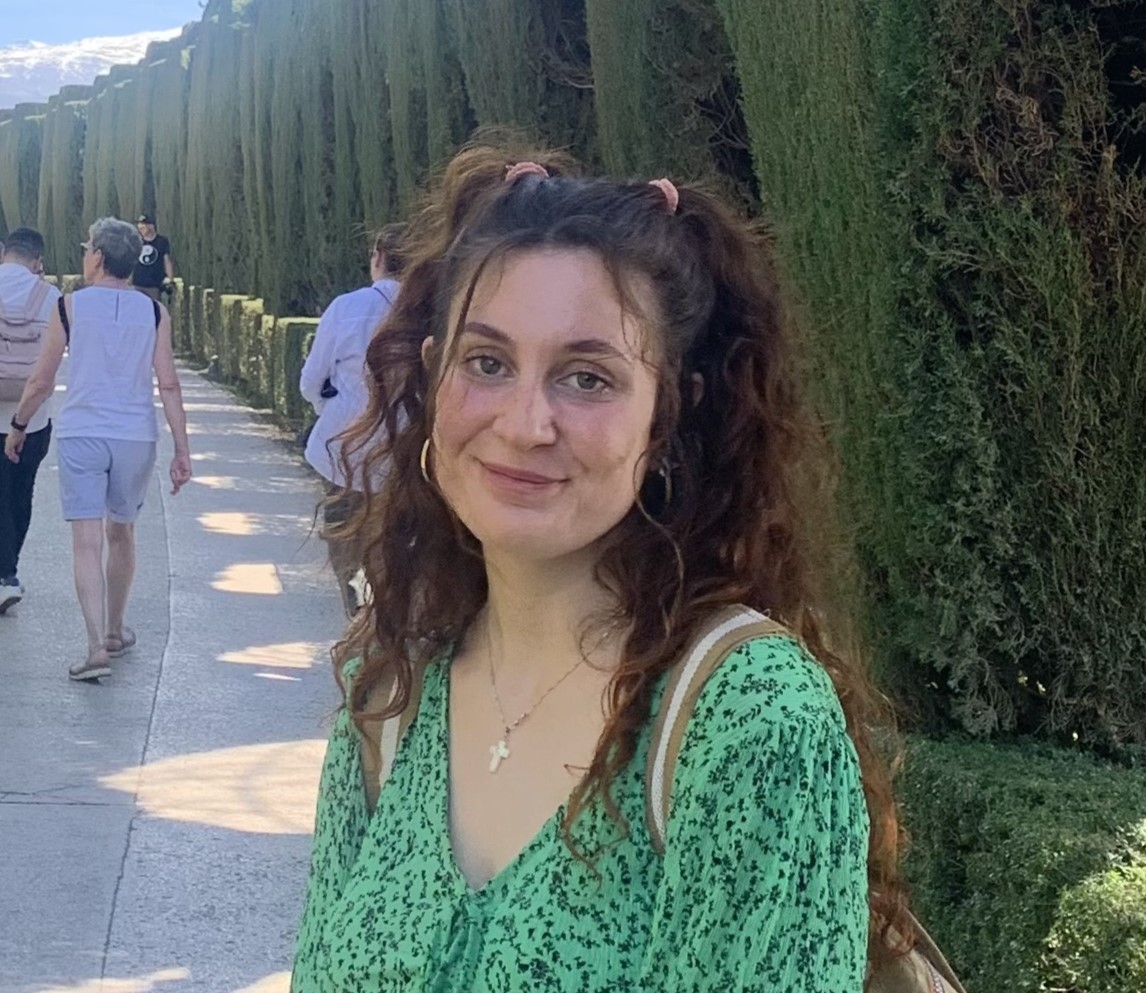}}]{Blanca Lopez} received her B.Sc. degree in Physics from the Universidad de Sevilla in 2022, followed by an M.Sc. degree in Physics and Mathematics from the Universidad de Granada in 2023. She is currently a Ph.D. candidate at IMDEA Networks and Universidad Carlos III de Madrid, where she contributes to the national MADQuantum-CM project and the international QSNP project as part of her doctoral research.
\end{IEEEbiography}
\begin{IEEEbiography}[{\includegraphics[width=1in,height=1.25in,clip,keepaspectratio]{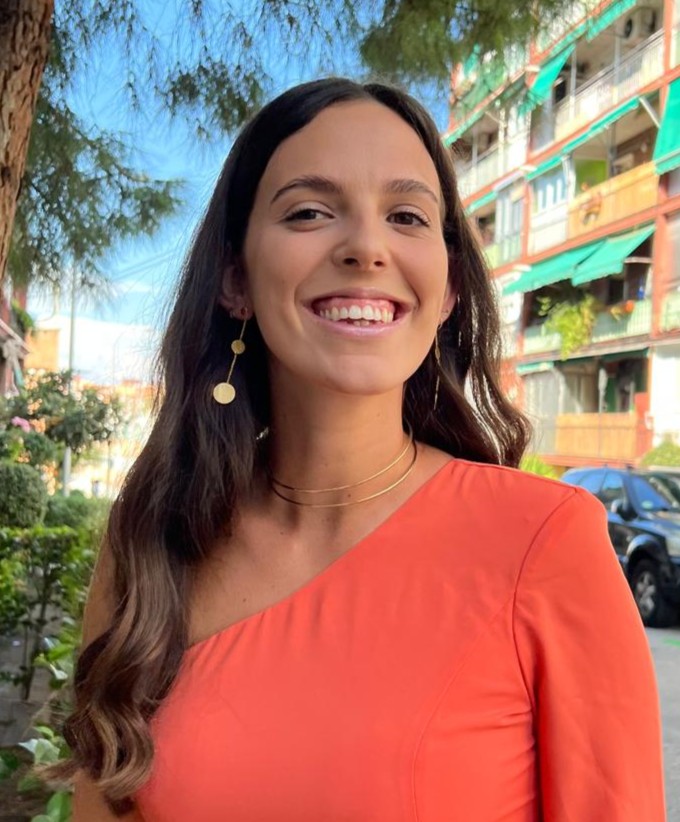}}]{Angela Diaz-Bricio} received a dual B.Sc. degree in Physics and Mathematics from Universidad Complutense de Madrid in 2023 and completed her Master’s in Quantum Technologies and Engineering at Universidad Carlos III de Madrid in early 2025. She began her PhD in March 2025 at IMDEA Networks and Universidad Carlos III de Madrid, where she is currently working on quantum communication systems as part of the MADQuantum-CM and QSNP projects.
\end{IEEEbiography}
\begin{IEEEbiography}[{\includegraphics[width=1in,height=1.25in,clip,keepaspectratio]{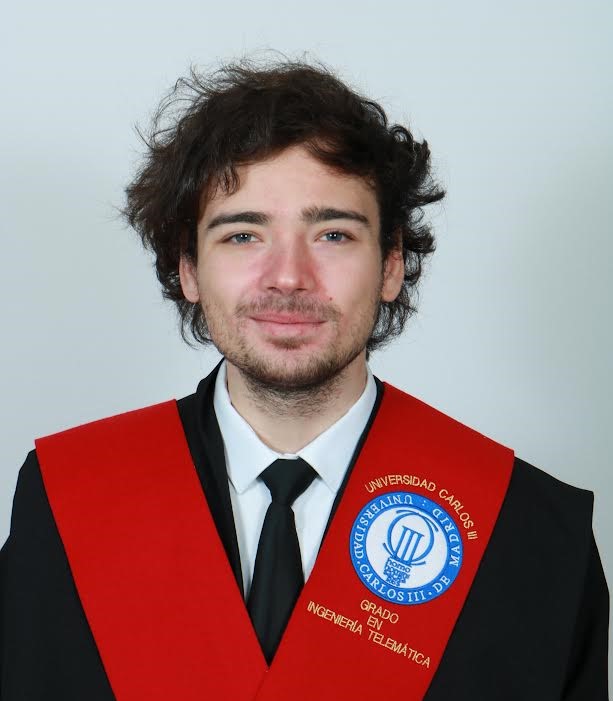}}]{Javier Perez} received his degree in Telematic Engineering from Universidad Carlos III de Madrid in 2024 and will begin a Master’s in Quantum Technologies and Engineering at Universidad Carlos III de Madrid in September 2025. He began his work as a Research Engineer in September 2025 at IMDEA Networks and Universidad Carlos III de Madrid as part of the MADQuantum-CM project.
\end{IEEEbiography}

\begin{IEEEbiography}[{\includegraphics[width=1in,height=1.25in,clip,keepaspectratio]{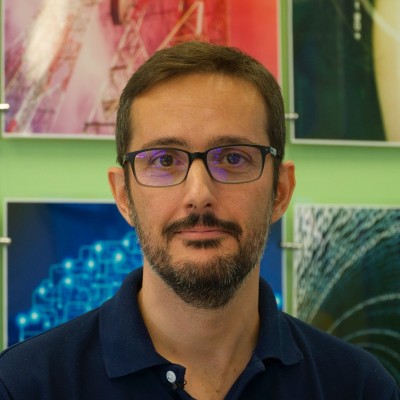}}]{Ivan Vidal}
received the Ph.D. in Telematics Engineering in 2008 from UC3M, where he works as Tenured Associate Professor. He has been involved in several international and national research projects, including the Horizon Europe NEMO, the TRUE5G, the QUBIP, and the MADQuantum-CM, and has published more than 60 scientific papers in several conferences and international journals.
\end{IEEEbiography}
\begin{IEEEbiography}[{\includegraphics[width=1in,height=1.25in,clip,keepaspectratio]{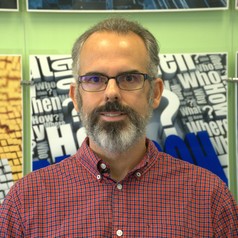}}]{Francisco Valera}
received the Ph.D. degree in telecommunications engineering from the University Carlos III of Madrid in 2002 where he is currently a Full Professor. He has published +100 scientific articles in the field of advanced communications in magazines and conferences. He has been involved in many research projects (HE QUBIP, HE PQREACT, QFE QSNP, MADQuantum-CM, etc.).
\end{IEEEbiography}
\vfill

\end{document}